# Coarse-Grained Analysis of Microscopic Neuronal Simulators on Networks: Bifurcation and Rare-events computations

**Konstantinos G. Spiliotis and Constantinos I. Siettos***

School of Applied Mathematics & Physical Sciences
National Technical University of Athens
Athens, GR-157 80, Greece

**Abstract.** We show how the Equation-Free approach for mutliscale computations can be exploited to extract, in a computational strict and systematic way the emergent dynamical attributes, from detailed large-scale microscopic stochastic models, of neurons that interact on complex networks. In particular we show how the Equation-Free approach can be exploited to perform system-level tasks such as bifurcation, stability analysis and estimation of mean appearance times of rare events, bypassing the need for obtaining analytical approximations, providing an "on-demand" model reduction. Using the detailed simulator as a black-box timestepper, we compute the coarse-grained equilibrium bifurcation diagrams, examine the stability of the solution branches and perform a rare-events analysis with respect to certain characteristics of the underlying network topology such as the connectivity degree.

**Keywords: Coarse-grained Computations, Complex Systems, Equation Free methodology, Multi-scale Dynamics, Neuronal networks, Bifurcation analysis, Nonlinear dynamics**

## 1. Introduction

Complex networks pertain to the structure of many real-world systems shaping their dynamics. It is not a surprise that over the last few years there has been an immense interest in studying the emergent complex dynamics (e.g. social, opinion, epidemic, neuronal) on networks. Due to the strongly heterogeneous nature of the topology of realistic networks as well as the stochastic and nonlinear very-large scale interactions of the underlying units, the emergent–macroscopic - behavior is most of the times far from trivial to predict. Self-organization, sustained oscillations, travelling waves, multiplicity of stationary states and spatio-temporal chaos are paradigms of the rich nonlinear behavior at the coarse-grained systems level. The brain with about $10^{12}$ neurons and $10^{15}$ neurons interconnections, packing and folding in 1.5kgr of mass, is the most complex system. The emergent cognitive dynamics are characterized by unique spatiotemporal patterns and functionalities [19].

One of the keys in order to model and study the brain behavior (in different anatomic regions such us the hippocampus, the thalamus and the cortex) is the network theory [41]. Lago – Fernandez et al., [23] studied the differences in the oscillatory behavior of neurons with respect the several types of networks (regular, Watts and Strogatz [45] and random networks). In order to explain the pathological situation in Parkinson Disease, Terman et al., [42] studied several types of networks (sparse normal and dense) between globus pallidus, external and subthalamic nucleus. Roxin et al., [34] studied networks with different densities of directed random connections and found that a localized transient stimulus results either in self-sustained persistent activity or in a brief transient behavior followed by failure. Studying the connection of epilepsy and the triggering properties of two areas of hippocampus CA3 regions (exhibiting short bursts of activity) and the CA1 regions (exhibiting seizure like activity lasting for seconds), Netoff et al., [29] constructed small-world network models of various types of neurons (Poisson spike train, leaky integrate-and-fire, stochastic Hodgkin and Huxley). For increasing values of the rewiring probability, the models were able to predict normal-like behavior, seizure-like transients and continuous bursting, i.e. phenomena observed in real situations.

For the systematic analysis and deeper understanding of such complex responses, one of the most critical issues in contemporary neuroscience, revolves around the bridging of the different scales: the micro-scale, where the interactions of the neurons take place and that of the macroscopic scale where the coarse-grained dynamic behavior emerges.

Traditionally, the gap between the high (microscopic) and the low-dimensional (macroscopic/coarse-grained) level is bridged through closures, relating higher-order, "fast" moments of the evolving detailed distributions to a few, low-order, "slow", "master" moments of the underlying detailed distributions [44, 4, 5, 6, 13, 31]. However, no general systematic methodology for deriving such closures exists; these are often based on assumptions that introduce certain biases in the analysis. Homogeneous neurons and network topology characteristics, infinite populations are some of the assumptions that may distort the qualitative and quantitative analysis at the coarse-grained level (see for example the critical discussion on the influence of certain closure approximations in the analysis of networks of neurons in [25]).

*Corresponding author; e-mail: ksiet@mail.ntua.gr

In the lack of good closures what is usually done is brute-force temporal simulation: one sets up different realizations of the underlying microscopic distributions, changes the values of the model parameters, possibly creates networks with different topologies and then runs the various scenarios for a long time; finally a statistical analysis of the emergent dynamics is performed to get the desired macroscopic observables. However this approach is inadequate for many important macroscopic-level computations: e.g. branches of unsteady fixed points or periodic oscillations cannot be detected in this way.

In this paper, we show how the so-called "Equation-free" approach [11, 26, 35, 17, 18, 38, 12, 27, 28, 36, 24, 40] a computational framework for multi-scale computations, can be used to efficiently extract "systems-level" information from detailed individualistic based neuronal- models with respect to certain topology characteristics of the underlying networks.

In particular, we show how the Equation-Free approach is exploited to perform system-level tasks such as bifurcation, stability analysis and estimation of mean appearance times of rare events, bypassing the need for obtaining analytical approximations, providing thus an "on-demand" model reduction. Using the detailed simulator as a black-box timestepper, we compute the coarse-grained equilibrium bifurcation diagrams, examine the stability of the solution branches with respect to certain characteristics of the underlying network topology such as the connectivity degree and perform a rare-events analysis.

For illustration purposes, we use a simple stochastic neuronal model based on the work presented in [22, 40]. The relatively simple rules governing the interactions of the neurons result in the emergence of complex dynamics in the coarse-grained level including phase transition from low to high densities of excited neurons and multiplicity of coarse-grained stationary states. Such behaviours have been observed in many brain activities [9]. Each neuron can be in one of two states: excited (activated) or non-excited (inactivated). Neurons interact with their neighbours on Erdős – Rényi networks [1, 3, 8, 30] in a undirectional way and change their states over discrete time in a probabilistic manner. Firstly, we study the dynamics of the network with respect to the activation probability and construct the coarse-grained bifurcation diagrams. We also construct the coarse-grained bifurcation diagrams with respect to the connectivity density of the networks. To our knowledge this is the first time that such systems-level tasks are derived *in an explicit manner* for interacting neuron populations on networks with respect to their topology characteristics.

The paper is organized as follows: in section 2, we describe the basic concept of the Equation-Free framework; in section 3 we give a description of the neuronal model; in section 4 we present and discuss the simulation results. We conclude in section 5.

**2. The Equation Free methodology for handling complexity: a multiscale computational approach**

**2.1 The basic concept.**

The Equation Free approach is a mathematical-assisted computational framework that can enable detailed microscopic simulators (in our case random networks of interacting neurons) to perform coarse-grained/ system-level tasks such as the construction of bifurcation diagrams, stability analysis, detection of coarse grained bifurcation points etc [17, 26, 35, 38, 39, 40]. Let us give a very short description of the methodology.

Let $U(x)$ denote the distribution function over the set of microscopic coordinates $x$ on a network. The main assumption of the method is that a macroscopic coarse grain description (in the form of ordinary or partial differential equations) exists and closes in terms of relatively few macroscopic variables, say $u \in R^m$. Typically, these are the low-order moments of the underlying microscopically evolving distributions. The existence of a coarse-grained model implies that the higher order moments, say, $u_f \in R^n$, of the distribution $U(x)$ become relatively quickly over the time scales of interest, "slaved" to the few lower ones, $u$ (Fig. 1).

However, these functionals, due to complexity of the underlying multiscale evolution of the dynamics, are unavailable or extremely difficult to derive. In fact, the Equation Free methodology provides these closures "on demand"; relatively short bursts of fine scale simulations establish naturally this slaving relation [11, 26, 35, 17, 18, 38, 36, 28, 24, 32].

A fundamental building block of the Equation-free algorithms is the coarse timestepper. This is the way to obtain macroscopic input–output information from a microscopic simulator. Under this framework, the first step concerns the definition of the coarse-grained variables. For example in a neuronal network, the coarse variables can be the density of activated neurons or more generally the densities $d_k$, ($k = 1, 2, ..., k\max$) of the activated neurons with degree (number of connections) $k$. Once the appropriate macroscopic observables, have been indentified, the coarse time-stepper consists of the following essential components (see also Fig. 2).

1. Prescribe a coarse grained initial condition, say $\boldsymbol{u}(t_0)$. Define also a restriction operator $\boldsymbol{M}$ mapping the microscopic-level description to the macroscopic one; that is: $\boldsymbol{u} = \boldsymbol{M}\boldsymbol{U}$
2. Transform the macroscopic initial condition $\boldsymbol{u}(t_0)$ through a lifting operator $\boldsymbol{\mu}$, into consistent microscopic realizations: $\boldsymbol{U}(t_0) = \boldsymbol{\mu}\boldsymbol{u}(t_0)$.
3. Evolve these realizations in time using the microscopic simulator for a short macroscopic time $T$, generating the microscopic distribution $\boldsymbol{U}(t_0 + T)$. The choice of $T$ is associated with the (estimated) spectral gap of the linearization of the unavailable closed macroscopic equations [15, 33].
4. Obtain the coarse grained variables using the restriction operator $\boldsymbol{M}$: $\boldsymbol{u}(t_0 + T) = \boldsymbol{M}\boldsymbol{U}(t_0 + T)$. Lifting from the macroscopic to microscopic and then restricting again to macroscopic should not have an effect except round off errors, which means that $\boldsymbol{I} = \boldsymbol{M}\boldsymbol{\mu}$.

The above procedure can be considered as a map from the initial macroscopic conditions $(\boldsymbol{u}(t_0), \boldsymbol{p})$ (where $\boldsymbol{p} \in R^q$ is the parameter vector of the system), to the result of the time integration after a given time-horizon $T$, i.e.

$$\boldsymbol{u}(t_0 + T) = \boldsymbol{\Phi}_T(\boldsymbol{u}(t_0), \boldsymbol{p}) \tag{1}$$

At this point one can utilize computational superstructures like the Newton's or Newton- GMRES method [15, 16] as a shell "wrapped around" the coarse timestepper (1) to compute and trace branches of coarse-grained equilibria or periodic solutions (even unstable ones) [14]; other computational superstructures (other "wrappers") e.g. iterative eigensolvers such as Arnoldi's algorithm [14, 37] can also be used to extract information about the stability of the coarse-grained system dynamics.

For example coarse-grained (macroscopic) equilibria can be obtained as fixed points of the map :

$$\boldsymbol{u} - \boldsymbol{\Phi}_T(\boldsymbol{u}, \boldsymbol{p}) = 0 \tag{2}$$

To trace fixed point branches through turning points, in one parameter space, one can augment the fixed point equation (2) with the linearized pseudo arc-length continuation condition

$$N(\boldsymbol{u}, s) = \boldsymbol{a}(\boldsymbol{u} - \boldsymbol{u_0}) + \beta(p - p_0) - \Delta s = 0 \tag{3}$$

$\Delta s = s - s_0$ is the pseudo arc-length continuation step and

$$\boldsymbol{a} = \frac{\boldsymbol{u_1} - \boldsymbol{u_0}}{\Delta s} \text{ and } \beta = \frac{p_1 - p_0}{\Delta s}.$$

$(\boldsymbol{u_0}, p_0)$ and $(\boldsymbol{u_1}, p_1)$ are two already known solutions of (2) which can be found for example by temporal simulations. The tracing of the solution branches can be obtained using an iterative procedure like the Newton-Raphson technique. The procedure involves the iterative solution of the following linearized system:

$$\begin{bmatrix} I - \dfrac{\partial \boldsymbol{\Phi}_T}{\partial \boldsymbol{u}} & -\dfrac{\partial \boldsymbol{\Phi}_T}{\partial p} \\ \boldsymbol{a} & \beta \end{bmatrix} \begin{bmatrix} \delta \boldsymbol{u} \\ \delta p \end{bmatrix} = -\begin{bmatrix} \boldsymbol{u} - \boldsymbol{\Phi}_T(\boldsymbol{u}, p) \\ N(\boldsymbol{u}, p) \end{bmatrix} \quad (4)$$

Note that for the calculation of the Jacobian $\dfrac{\partial \boldsymbol{\Phi}_T}{\partial \boldsymbol{u}}$ and $\dfrac{\partial \boldsymbol{\Phi}_T}{\partial p}$, no explicit macroscopic evaluation equation are needed. They can be approximated numerically by calling the black-box coarse timestepper at appropriately perturbed values of the corresponding unknowns $(\boldsymbol{u}, p)$. The above framework enables the microscopic simulator to converge to both coarse-grained stable and unstable solutions and trace their locations [39, 14].

The above approach turns to be computationally inefficient for large scale problems. An alternative efficient approach in order to solve (2), (3) is the usage of matrix-free iterative solvers, such as the Newton-Generalized Minimum Residual (Newton-GMRES) method [15] . The advantage of using such matrix-free iterative solvers is that the explicit calculation and storage of the Jacobian is not required. Only matrix-vector multiplications are needed. These can be performed at low cost by calling the timestepper from nearby initial conditions allowing the estimation of the action of the linearization of a map $\boldsymbol{\Phi}_T$ on known vectors, since

$$D\boldsymbol{\Phi}_{T_h}(\boldsymbol{u}) \cdot \boldsymbol{q} \approx \dfrac{\boldsymbol{\Phi}_T(\boldsymbol{u} + \varepsilon \boldsymbol{q}) - \boldsymbol{\Phi}_T(\boldsymbol{u})}{\varepsilon} \quad (5)$$

where $\varepsilon$ is a small scalar. At each step $j$ GMRES minimizes the residual

$$R = \begin{pmatrix} R_1 \\ R_2 \end{pmatrix} = \begin{pmatrix} \boldsymbol{u} - \boldsymbol{\Phi}_T(\boldsymbol{u}) \\ N(\boldsymbol{u}, p) \end{pmatrix} \quad (6)$$

This is achieved by producing an orthonormal basis $\{q_1, q_2, ..., q_j\}$ of the Krylov subspace $K_j$ where $K_j = \{q_1, Jq_1, J^2 q_1, ..., J^j q_1\}$ and $J = \begin{bmatrix} I - \dfrac{\partial \boldsymbol{\Phi}_T}{\partial \boldsymbol{u}} & \dfrac{\partial \boldsymbol{\Phi}_T}{\partial p} \\ \boldsymbol{a} & \beta \end{bmatrix}$.

The projection of $J$ on $K_j$ is represented in the basis $V_j$ by the upper Hessenberg matrix $H_j = V_j^T J V_j$ whose elements are the coefficients $h_{ij}$ (see Fig. (3)).

### 2.2. Lifting and resting operators on discrete-state neuronal networks

Generally speaking, a network is described as a graph, i.e. a pair of set $G = (V, E)$, where $V$ is the set of vertices (or, as otherwise called, nodes) and $E$ is the set of the edges. A graph can be completely described by an adjacency matrix $A$ whose elements are defined as follows: if there is an edge between the nodes $i$ and $j$ then $A_{ij} = 1$; otherwise $A_{ij} = 0$. If the network is undirectional (which means that node $i$ is connected with node $j$ and vice versa) then $A_{ij} = A_{ji}$. The number of edges starting (or leaving) from node $i$ is called "degree of the node", say $k_i$. Regarding neuronal networks, the nodes are the neurons and the edges are the connections between them (chemical or electrical e.t.c). The set $V(k)$ of all neurons with degree $k$ is $V(k) = \{i : i \text{ has degree } k\}$ where $k = 1, 2, ... k_{max}$ and $k_{max} = \max(\deg rees)$, so, $\bigcup_{k=1,...k_{max}} V(k) = G$ and $V(k) \bigcap V(m) = \varnothing$ for $k \neq m$.

In a discrete-state neuronal network, the state of the $i$-th cell at time $t$ is denoted by $x_i(t)$, where $x_i(t) \in \{0,1,2,...p\}$. Usually, the values $\{1,2,...p-1\}$ correspond to the refractory period; the value $x_i(t) = p$ corresponds to excitable state and the value $x_i(t) = 0$ is the firing state [43]. Here at each time step, we define as coarse macroscopic variables the densities of active neurons with degree $k$, $k = 1,2,...\max(\deg ree)$ i.e.:

$$d_k(t) = \frac{\sum_{j \in V(k)} a_{k,j}(t)}{N} \quad (7)$$

where $a_{k,j}(t) = 1$ if $x_i(t) = 1$ meaning that the $i$-th neuron with degree $k$ is fired. Also, let $\rho(t)$ be the total density of activated neurons i.e. the number of active neurons divided by the total number of neurons in the network i.e.

$$\rho(t) = \frac{\sum_{i \in G} a_i(t)}{N} \quad (8)$$

Then the $l_1$ norm of the vector of densities $\boldsymbol{d} = (d_1, d_2, ..., d_{k\max})$ reads:

$$\|\boldsymbol{d}\|_1 = |d_1| + |d_2| + ... + |d_k| = d_1 + d_2 + ... + d_{k\max} = \frac{\sum_{k=1}^{k\max} \sum_{j \in V(k)} a_{k,j}}{N} = \frac{\sum_{i \in G} a_i}{N} = \rho \quad (9)$$

The restriction operator is a map $\boldsymbol{M}$ from $N_{copies}$ different realizations of the state matrix of neurons, $\left[\{0,1\}^N\right]^{Ncopies}$ ($N$ the number of neurons), to the vector of densities of the degrees taken by averaging the $N_{copies}$ realizations i.e.

$$\boldsymbol{M} : \left[\{0,1,2,...n\}^N\right]^{Ncopies} \to R^{k\max} \text{ and } \boldsymbol{d} = (d_1, d_2, ..., d_{k\max}) = \boldsymbol{MU} \quad (10)$$

The lifting operator $\boldsymbol{\mu}$ transforms the vector of densities of degrees to appropriate $N_{copies}$ microscopic states i.e.

$$\boldsymbol{\mu} : R^{k\max} \to \left[\{0,1,2,...n\}^N\right]^{Ncopies} \text{ and } \boldsymbol{U} = \boldsymbol{\mu d} \quad (11)$$

**2.3 Rare-events analysis**

Consider a general one-dimensional stochastic process $\psi(t)$. The evolution of the probability density $P(\psi, t)$ obeys the following master equation [33, 10, 21]:

$$P(\psi, t) = \int p(\psi, t+\tau | \psi', t) P(\psi', t) d\psi' \quad (12)$$

where $p(\psi, t+\tau | \psi', t)$ is the transition probability from point $\psi'$ at time $t$ to point $\psi$ at time $(t+\tau)$. The differential form of this equation, known as the Kramers-Moyal expansion, reads:

$$\frac{\partial P(\psi, t)}{\partial t} = \sum_{n=1}^{\infty} \left(-\frac{\partial}{\partial \psi}\right)^n D^{(n)}(\psi, t) P(\psi, t) \quad (13)$$

where

$$D^{(n)}(\psi,t) = \frac{1}{n!} \lim_{\tau \to 0} \frac{1}{\tau} \left\langle \left(\xi(t+\tau) - \xi(t)\right)^n \right\rangle \bigg|_{\xi(t)=\psi} \tag{14}$$

are the differential moments of the transition probability $p$ [10, 33]. The angular brackets here denote ensemble averaging and $\xi$ denotes a realization of the stochastic process with a δ-function distribution at the starting point $\xi(t) = \psi$. Under certain assumptions [21, 33] the equation (14) takes the form.

$$\frac{\partial P(\psi,t)}{\partial t} = \left[ -\frac{\partial}{\partial \psi} u(\psi) + \frac{\partial^2}{\partial \psi^2} D(\psi) \right] P(\psi) \tag{15}$$

Here, $u(\psi) = D^{(1)}(\psi)$ and $D(\psi) = D^{(2)}(\psi)$ are the drift and the diffusion coefficients respectively, which can be estimated by the formula (14) which can be written as

$$u(\psi) = \lim_{\tau \to 0} \frac{\langle \Delta\psi(t,\psi) \rangle}{\tau} \tag{16}$$

$$2D(\psi) = \lim_{\tau \to 0} \frac{\langle \Delta\psi(t,\psi)^2 \rangle}{\tau} \tag{17}$$

where

$$\Delta\psi(t,\psi) = \psi(t+\tau, \psi) - \psi(t,\psi) \tag{18}$$

The Fokker-Planck equation components (the drift and the diffusion coefficient) can be estimated by applying the Equation Free framework. Specifically,

a. We initialize the system consistently with the coarse variable $\psi$.
b. Through the lifting operator $\mu$, transform the macroscopic initial condition $\psi$ into consistent microscopic realizations: $U(t) = \mu\psi(t)$.
c. Use the microscopic simulator for a short macroscopic time $\tau$ to get $U(t+\tau)$.
d. Obtain the coarse grained variables using the restriction operator $M$: $\psi(t+\tau) = MU(t+\tau)$.
e. Set the value $\psi(t+\tau)$ in (18) and calculate the drift and the diffusion coefficient (16) and (17).

Once the Fokker-Planck equation is reconstructed, one can calculate several global characteristics of the system, such as the effective free energy $G(\psi)$ and the rates of transitions between different metastable states of the system [7, 12, 21, 10, 33, 38] using the following realtion:

$$\beta G(\psi) = -\int_0^\psi \frac{u(\psi')}{D(\psi')} d\psi' + \ln D(\psi) + const \tag{19}$$

When the potential barrier between the stable and the unstable point is relatively big to the noise, the mean escape rates between two metastable coarse-grained states can be estimated as [7, 21]:

$$\tau \approx \int_{\psi_{stable}}^{\psi_{unstable}} \exp\left[\beta G(\psi)\right] d\psi \int_{-\infty}^{\psi_{unstable}} \frac{1}{D(\psi)} \exp\left[-\beta G(\psi)\right] d\psi \tag{20}$$

**3. The stochastic neuronal model**

The model considered here is a natural extension of the model presented in Kozma et al. [22] and Spiliotis and Siettos [40]. The model consists of $N$ neurons interacting on an Erdős-Rényi type of network [8, 1, 3, 30]. The construction of such a network is done according to the following simple rule: each neuron is connected with another neuron with probability $p$. The resulting degree distribution is a binomial distribution [1, 3, 30]:

$$P(k) = \binom{N-1}{k} p^k (1-p)^{N-1-k} \qquad (21)$$

where $\binom{N-1}{k}$ are the combinations $N-1$ to $k$. As $N \to \infty$, the binomial distribution can be approximated by the Poisson distribution,

$$P(k) = \frac{e^{-\bar{k}} (\bar{k})^k}{k!} \qquad (22)$$

In Fig.4, we give an example with $N = 10000$ nodes-neurons and connectivity probability $p = 0.0008$. The distribution is symmetric around the mean value, which is $\bar{k} = pN$. If $\bar{k} < 1$ the network has many isolated trees. If $\bar{k} > 1$ there are a giant cluster in the graph and for $\bar{k} > \ln N$ almost every graph is totally connected [1, 3, 30]. The mean length is proportional to

$$l \sim \frac{\ln N}{\ln \bar{k}}$$

Finally, two neighborhoods of a node-neuron are also connected, with probability $p$. Hence, the clustering coefficient is given by

$$c_i = \frac{2E_i}{k_i(k_i-1)} = p \qquad (23)$$

where $E_i$ is the number of triangles corresponding to the $i-th$ neuron.

Each neuron is labeled as $i = 1,...,N$ and takes two values: the value "1" if is it activated and the value "0" if it is not. We describe the state of the $i-th$ neuron in time $t$ with the function $a_i(t) \in \{0,1\}$. Let $\Lambda(i)$ be the set of the neighbors (i.e. the neurons connected to $i-th$ neuron). Also consider the summation

$$\sigma_i(t) = \sum_{j \in \Lambda(i)} a_j(t) \qquad (24)$$

which gives the number of activated neighbors of the $i$ neuron. At each time step each neuron interacts with its neighboring neurons, and changes its state-value according to the following stochastic way:

1. An inactive neuron becomes activated with probability $\varepsilon$, if $\sigma_i(t) \leq k_i$ ($k_i$ is the degree of the $i-th$ neuron) and at the same time there is at least an active neighboring neuron. If $\sigma_i(t) > k_i$ the neuron becomes activated with probability $1-\varepsilon$.
2. An activated neuron becomes inactive with probability $\varepsilon$, if the $\sigma_i(t) > k_i$. If $\sigma_i(t) \leq k_i$ the neuron becomes inactivate with probability $1-\varepsilon$.

The range of the parameter $\varepsilon$ is the interval $(0, 0.5)$. The above rules are usually called majority rules, due to the way they operate [22]. For example if the majority of neighbors are active, i.e. $\sigma_i(t) > k_i$, then the probability of an inactive neuron becoming active is $1 - \varepsilon$ (which is big because $\varepsilon \ll 1$) which means that the inactive neuron follows the majority. We give two examples of the above rules in Fig. 5.

## 4. Numerical results

### 4.1 Analysis with respect the activation probability $\varepsilon$

The results are obtained using networks of $N = 10000$ neurons. We performed a coarse-grained analysis on three different Erdős – Rényi networks. These were constructed with connectivity probabilities $p = 0.0006$  $p = 0.0007$ and $p = 0.0008$.

Fig. 6 depicts the time evolution of $\boldsymbol{d}$ for connectivity probability $p = 0.0006$. For small values of the parameter $\varepsilon$ and depending on the initial conditions, the system has two steady state solutions, one corresponding to zero densities, the "all-off state" (i.e. the all neurons turn inactive) and the other corresponding to high values of densities, meaning that the majority of the neurons are activated (Fig. 6a). As the value of $\varepsilon$ increases the "all-off" state solution loses its stability at a certain point where another non-zero solution, with small values of densities, springs (Fig. 6b). For higher values of $\varepsilon$ the high-densities solution disappears and now independently on the initial conditions the network converges to low densities states (the majority of the neurons turn inactive). These states are the only possible ones for even higher values of $\varepsilon$ (Fig. 6c, d)

Figs. 7 and 8 give the temporal evolution of $\boldsymbol{d}$ for connectivity probability $p = 0.0007$ and $p = 0.0008$ respectively. The dynamics are similar to the ones appearing in Fig. 6. For small values of the parameter $\varepsilon$, there are two stable steady states (the zero and the high-densities ones) (Fig. 7(a) and 8(a)). As the parameter $\varepsilon$ increases, the zero solution disappears and gives birth to a low densities state solution (Fig. 7(b),(c) and 8(b),(c)) which coexists with the high densities state. For higher values of $\varepsilon$, the high densities state solution diminishes (Fig. 7(b),(c) and 8(b),(c)) and finally disappears (see Fig. 7(d), 8(d)). The above results are indicative of the existence of several critical parameter values marking the onset of sudden qualitative changes in the dynamic behavior of the system. As the bifurcation theory suggests, this implies the emergence of unstable coarse-grained states which are unreachable through plain long-run temporal simulations.

The bifurcation diagrams with respect to the activation probability parameter $\varepsilon$ were constructed exploiting the Equation – Free framework as described in the previous sections. The coarse-grained variables are the densities $\boldsymbol{d} = (d_1, d_2, ..., d_{k\max})$ of the neurons with degree $k$. At time $t_0$, we created $N_{copies}$ different microscopic realizations (on the same network) consistent with the macroscopic variables $\boldsymbol{d}$. The coarse timestepper is constructed as the $T$ – map :

$$\boldsymbol{d}(t_0 + T) = \boldsymbol{\Phi}_T(\boldsymbol{d}(t_0), \varepsilon) \qquad (25)$$

The derived coarse-grained bifurcation diagrams for $p = 0.0008$, $p = 0.0007$, and $p = 0.0006$ are depicted in Fig. 9,11,13 respectively. The stationary states on the coarse-grained bifurcation diagram have been obtained as fixed points of (25) averaging over $N_{copies} = 20000$ realizations. Continuation around the coarse-grained turning points is accomplished by coupling the fixed point equation (25) with the pseudo-arc-length continuation, equation (3). The augmented linearized system (4) is solved by wrapping Newton-GMRES (section 2.1), as a shell, around the coarse timestepper. In these diagrams, solid lines correspond to stable coarse-grained states while dotted lines to unstable ones. The diagram consists of two "families" of solutions. One family of solutions is characterized by a saddle node bifurcation: the high densities state bifurcates through a turning point ($\varepsilon = 0.209$, $\varepsilon = 0.1889$ and $\varepsilon = 0.156$ for the networks that were constructed with probability $p = 0.0008$, $p = 0.0007$ and $p = 0.0006$, respectively) marking the change in the stability. Another familiy of lower densities bifurcates through a transcritical bifurcation. The "all-off" phase, which exists for all values of $\varepsilon$, is stable for small values of $\varepsilon$ and unstable beyond a transcritical point ($\varepsilon = 0.105$, $\varepsilon = 0.119$ and $\varepsilon = 0.134$, for $p = 0.0008$, $p = 0.0007$ and $p = 0.0006$) giving rise to a branch of a stable non-zero phase of relatively low network activities. The critical eigenvalues (see Figs. 10, 12, 14) are computed by implementing an Arnoldi iterative eigensolver which was again wrapped around our coarse timestepper [14, 16, 17, 37]. As it is shown, there is one eigenvalue with a bigger magnitude compared with the other five eigenvalues; past the turning point on the unstable branch, the magnitude of this leading eigenvalue is greater than one as expected. Figs. 10(c), 12(c), 14(c) depict the corresponding "maximum" eigenvalue with respect to $\varepsilon$. Figure 15 summarizes the obtained results for all three networks.

In Figure 16 we also show the bifurcation diagram obtained by a low-order mean field (MF) approximation for the total density $\rho$ of activated neurons, for a random network with constant connectivity degree equal to eight. The coarse-grained bifurcation diagram obtained using the detailed individual-based neuronal model on a Erdős – Rényi network with an equal mean value connectivity (as constructed by setting $p = 0.0008$) is also shown for comparison reasons. In the (MF) approximation the time evolution of total density is given by

$$\rho_{t+1} = f(\rho_t) \tag{26}$$

The function $f$ is being formulated [2, 22] by

$$f(\rho) = (1-\varepsilon) \sum_{i=0}^{\bar{k}/2-1} \binom{\bar{k}}{i} \rho^{\bar{k}-i}(1-\rho)^i + \varepsilon \sum_{i=\bar{k}/2}^{\bar{k}} \binom{\bar{k}}{i} \rho^{\bar{k}-i}(1-\rho)^i \tag{27}$$

where $\binom{\cdot}{\cdot}$ is the binomial coefficient. The bifurcation diagram comes up by solving the equation.

$$\rho = f(\rho) \tag{28}$$

The relatively simple mean field model assumes randomly selected neurons with replacement, omitting therefore spatial correlations, while the size of the population is considered infinite. Compared to the results obtained by the Equation-Free approach, the MF approximation gives a qualitatively similar bifurcation diagram, yet a quantitatively different one. In particular, close to the coarse-grained criticalities, the analytical model deviates from the actual detailed simulation results, while the Equation-Free framework captures the correct coarse-grained behaviour.

**4.2. Analysis with respect the connection probability $p$ (sparseness - denseness) of the networks.**

Figure 15 reveals that as the value of the connection probability connection decreases, the critical value of the bifurcation parameter also decreases. In order to examine how the structure of the network influences the dynamics behavior of the system, we keep the value of the parameter $\varepsilon$ constant, and we construct the coarse-grained bifurcation diagram with respect to the connection probability $p$ of the network. Here, the lifting operator generates $N_{copies}$ different networks, constructed with the same value of connectivity $p$, consistent with the initial macroscopic $d(t_0)$. Again, after time $T$, the restriction operator averages over the ensambles of the $N_{copies}$ realizations. The coarse timestepper reads

$$d(t_0 + T) = \boldsymbol{\Phi}_T(d(t_0), p) \tag{29}$$

Exploiting the proposed framework, we constructed the bifurcation diagram shown in Fig. 17. It's worth noting that the bifurcation diagram has been constructed in the set of all Erdős – Rényi networks, which describe the universality of the method. Starting from a value around $p = 0.0008$, the network has four solutions two of which are unstable. As the value of $p$ decreases (every value of $p$ corresponds to different networks), resulting to sparser network topologies, the norm of the high densities state decreases and at a critical point around $p^* = 0.549$ a branch of unstable coarse-grained states emerges. The zero (all-off state) solution always exists and it is stable for small values of $p$; at a critical value around $p^* = 0.612$ loses its stability through a transcritical bifurcation and gives rise to a branch of stable, yet low densities states. Figure 18 gives the six, larger in magnitude, eigenvalues of the high densities state around the turning point, on the stable (Fig. 18a) and on the unstable (Fig. 18b) branch.

**4.3. First time passage estimation**

Fig.19 depicts a long run time evolution of $\mathbf{d}$ for connectivity probability $p = 0.0007$ and $\varepsilon = 0.181$. As it is shown the coarse-grained steady states are metastable ones: after a relatively long period of time – which depends on the probability $\varepsilon$ - the system "jumps" from the high to the low densities state. The inverse jump is also possible.

A basic assumption that has to be fulfilled in order to perform the rare-event analysis as presented in section 2.4, is that the noise is white (i.e. its distribution is Gaussian). Fig.20 illustrates the probability density function of the noise as computed from a long-run temporal simulation around the stationary state at $\varepsilon = 0.181$. As it is shown, the resulting distribution can be well approximated by a Gaussian distribution with a mean value equal to $m = 0.72$ and a standard deviation $\sigma = 0.0123$. Here the reaction coordinate is defined as

$$\psi = (\mathbf{d}_{node} - \mathbf{d})^T \frac{(\mathbf{d}_{saddle} - \mathbf{d}_{node})}{\|(\mathbf{d}_{saddle} - \mathbf{d}_{node})\|} \tag{30}$$

corresponding to the projection of the difference of the coarse state from the coarse-grained node on the normalized difference between the coarse node and the saddle.

Fig. 21 shows the approximated coarse-grained free-energy with respect to $\psi$ as computed by the coarse timestepper, from short runs of $T = 7$ and 4000 realizations. Applying Kramer's theory we found a mean time to escape of the order of $2 \cdot 10^5$ time steps which comes in a good agreement with the long-run temporal simulations.

**5. Conclusions**

We have demonstrated how the Equation-Free approach, a computational framework for multi-scale computations, can be exploited to systematically analyze the coarse-grained dynamics of networks of neurons with respect to certain topological attributes. For illustration purposes, we used a simple individual-based stochastic model which, however, is able to catch a fundamental feature of such problems, which is the emergence of complex dynamics in the coarse-grained level including phase transitions and multiplicity of coarse-grained stationary states. The scheme allows the tracing of coarse-grained equilibria (stable and unstable), the investigation of their stability and the estimation of mean appearance times of rare events. This is achieved in a strict computationally way bypassing the need for obtaining analytical approximations, providing thus an "on-demand" model reduction. A comparison with a mean field approximation was also given, revealing the merits of the proposed approach. Two are the fundamental assumptions of the approach, namely: (a) time-scale separation between the micro and macro dynamics (b) the a-priori knowledge of the coarse-grained "slow" variables. Further research could proceed along several directions including the use of advanced data-mining techniques for such as Principal Component Analysis, or the Diffusion Maps [20] that can suggest the right coarse-grained observables as wells as the systematic analysis of more complex realistic network structures with applications to several neurological disorders such us Parkinson [42] or epilepsy diseases [29].

Figure 1

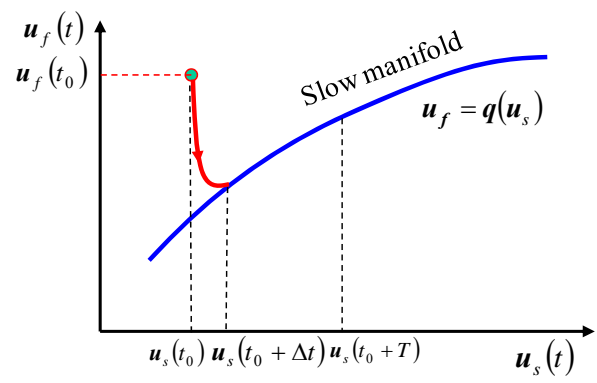

Figure 2

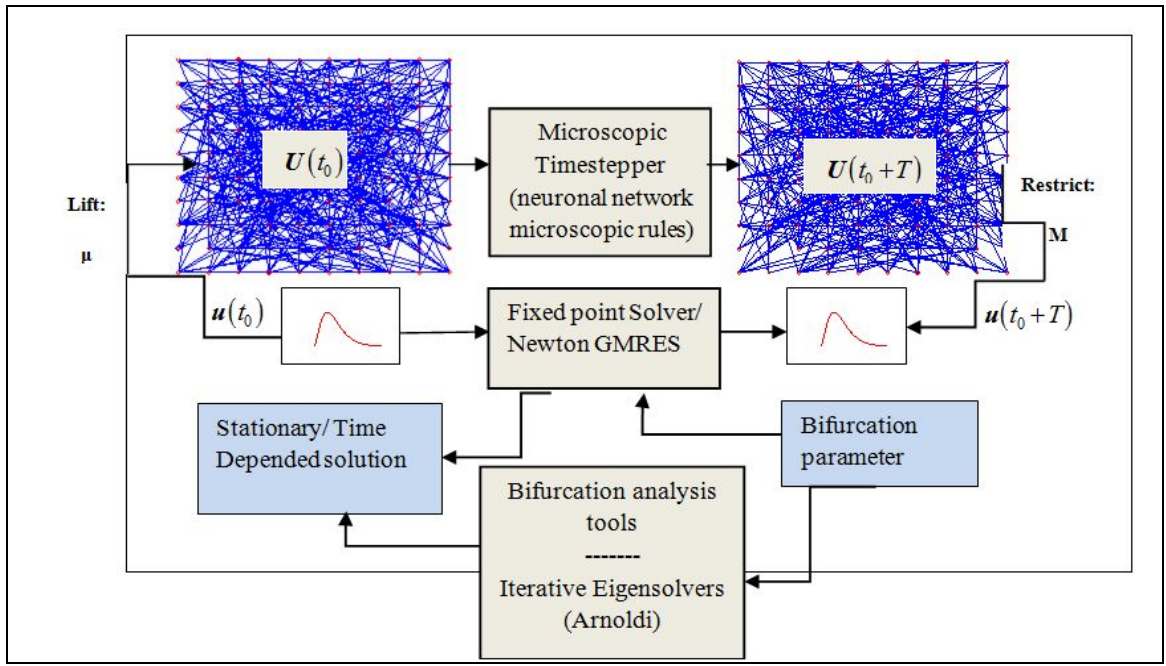

Figure 3

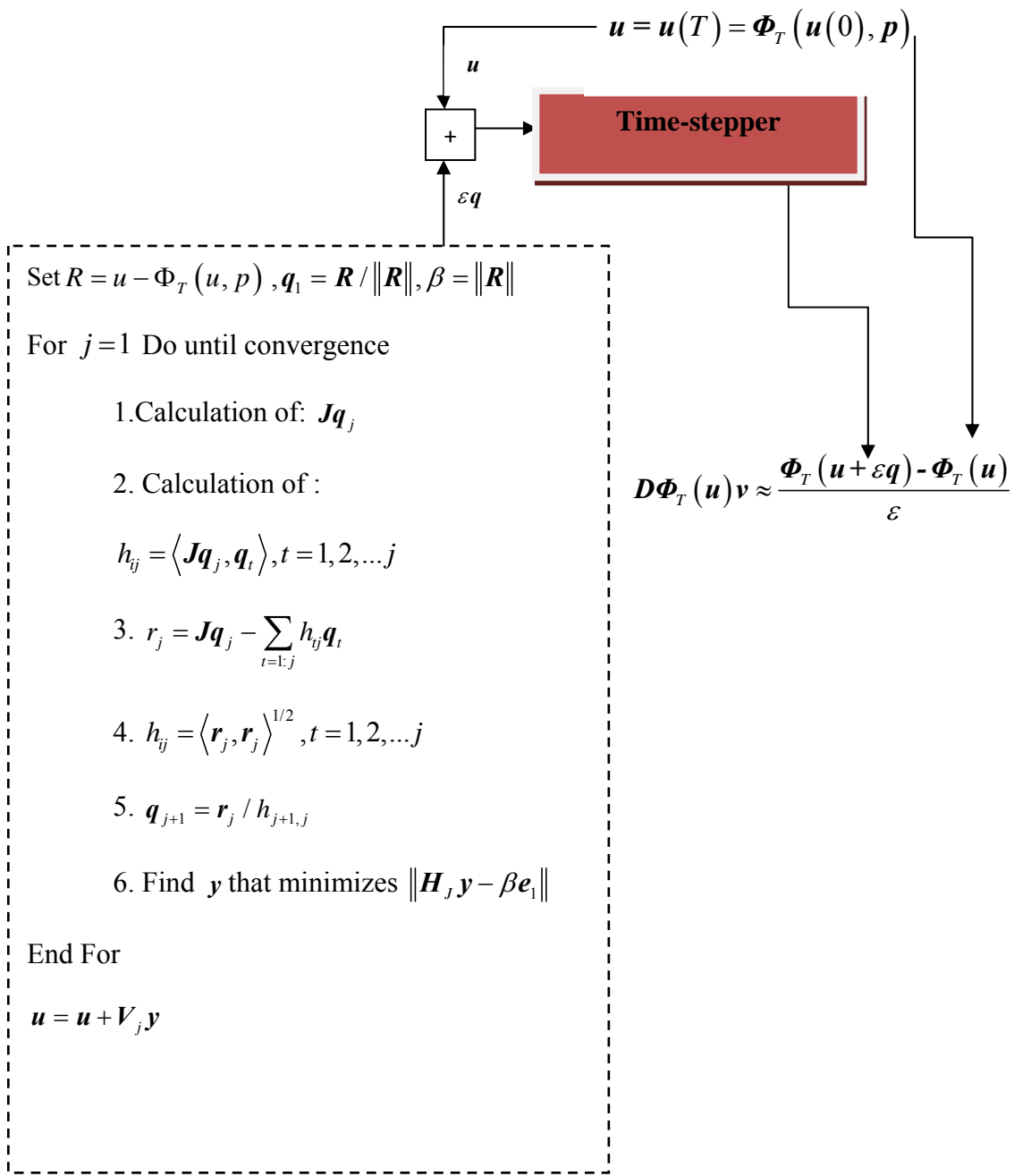

Figure 4

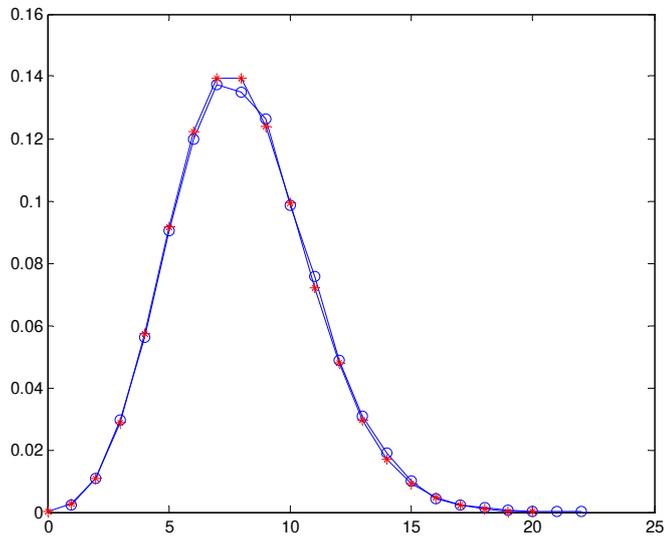

Figure 5

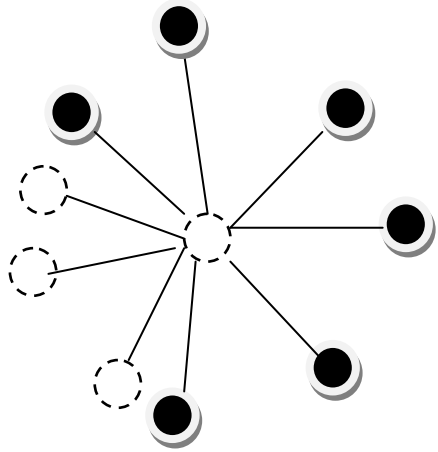 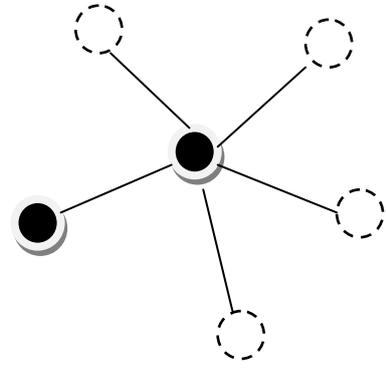

(a) (b)

Figure 6

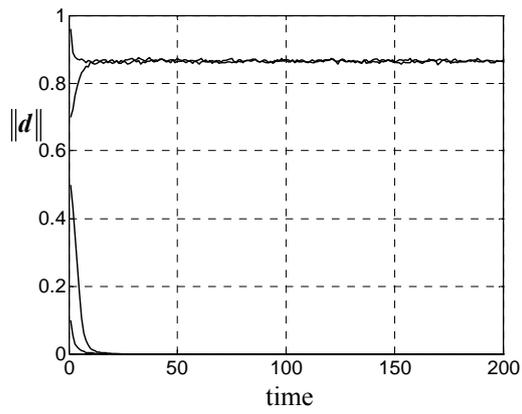

(a)

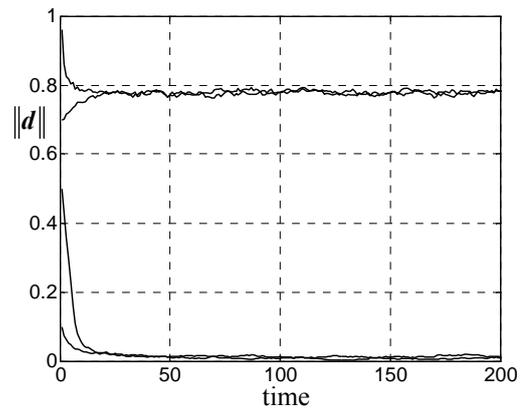

(b)

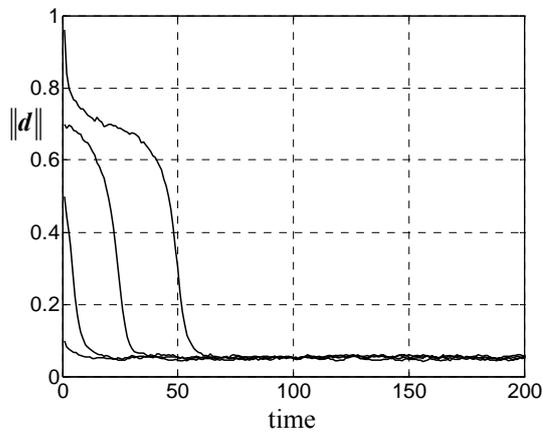

(c)

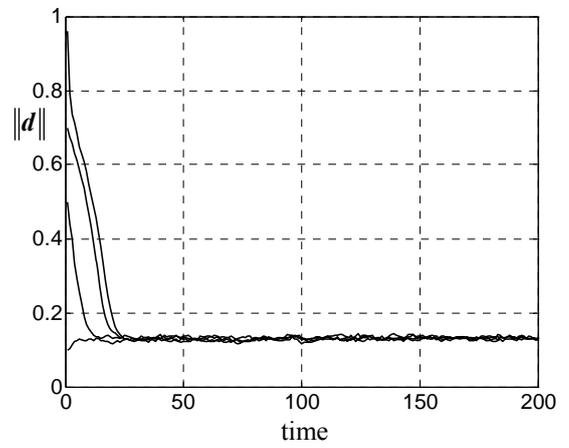

(d)

Figure 7

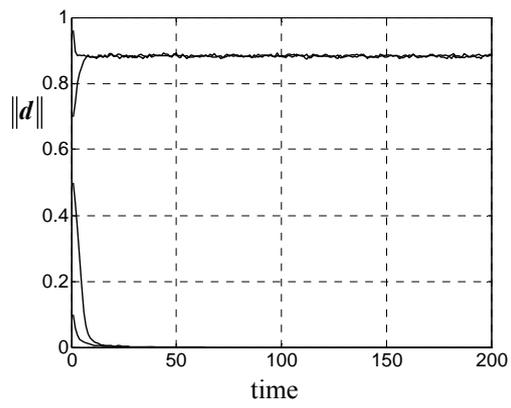

(a)

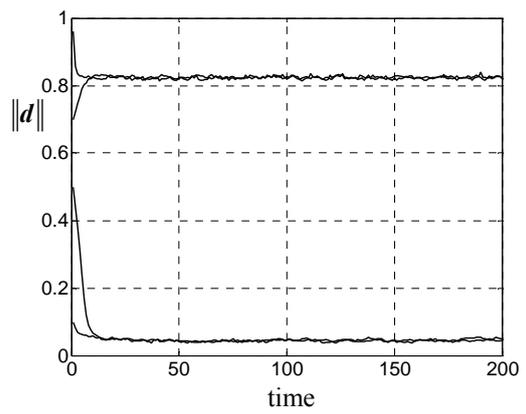

(b)

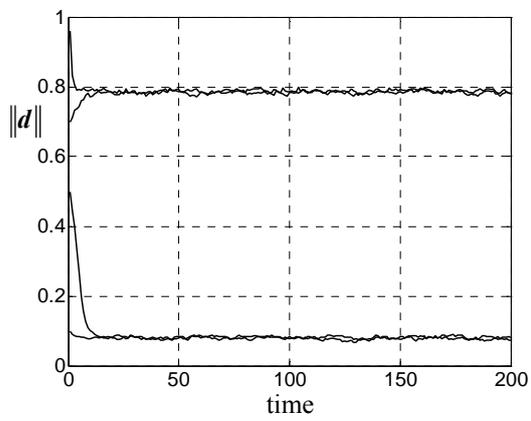

(c)

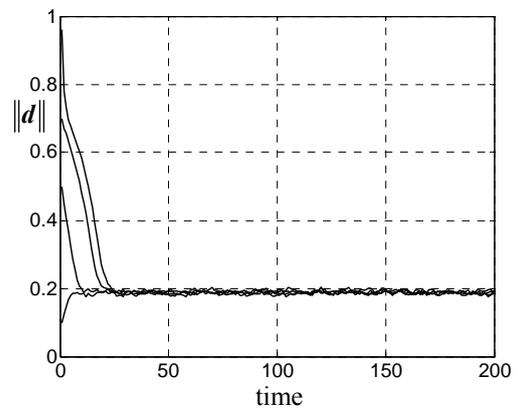

(d)

Figure 8

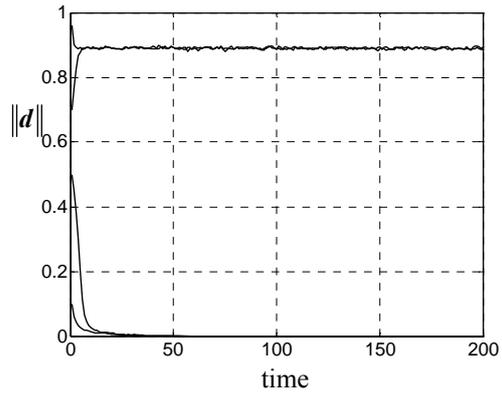

(a)

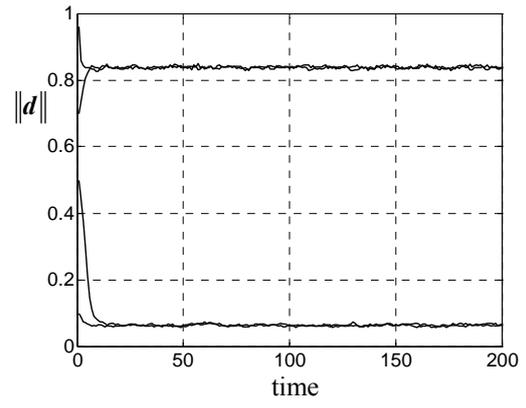

(b)

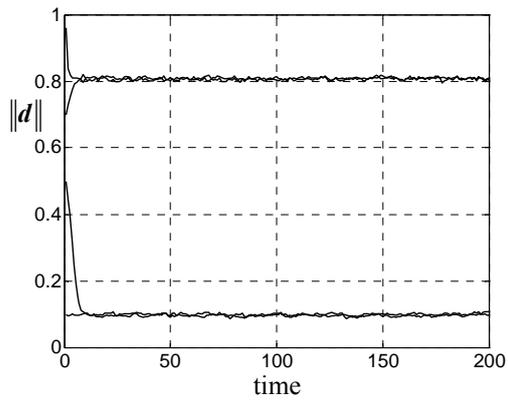

(c)

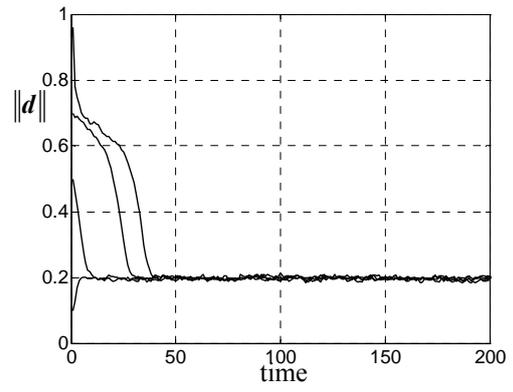

(d)

Figure 9

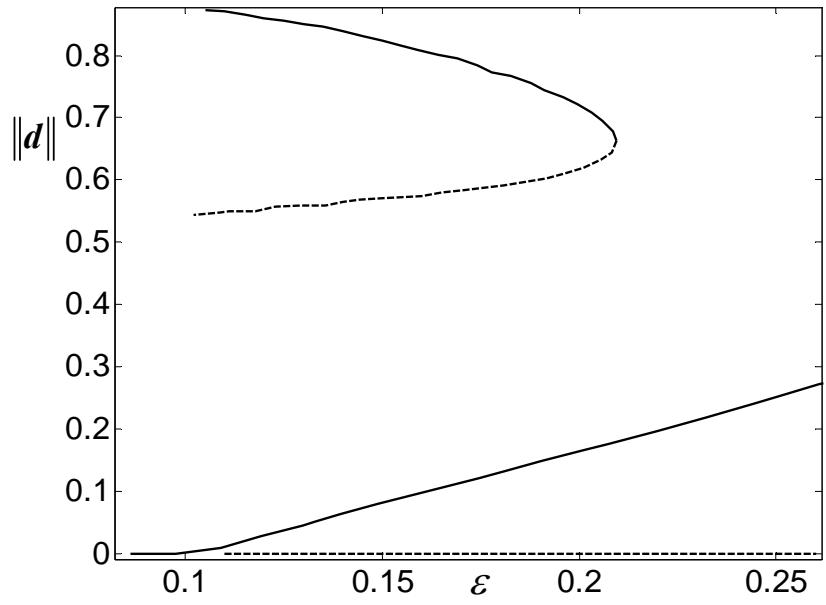

Figure 10

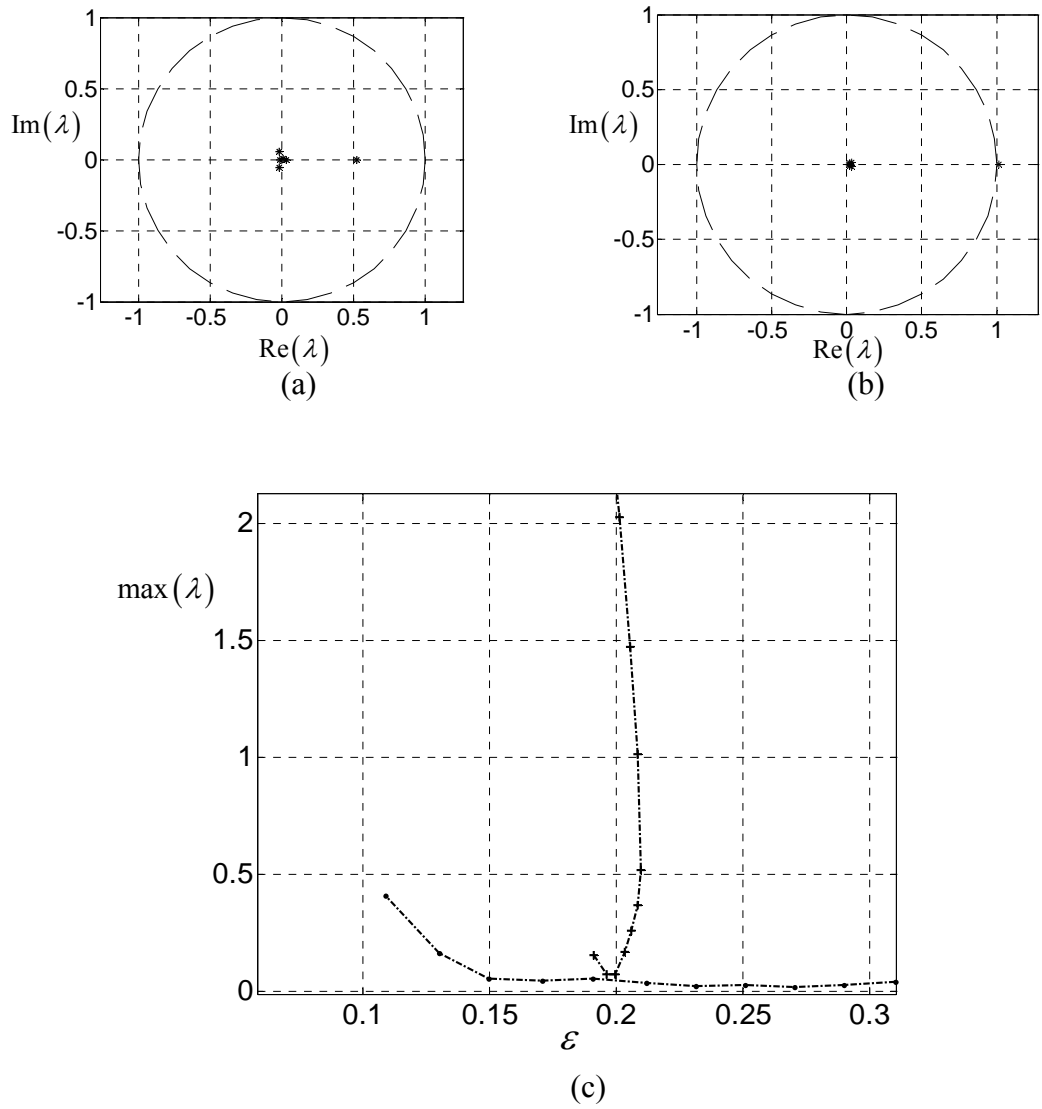

(a)

(b)

(c)

Figure 11.

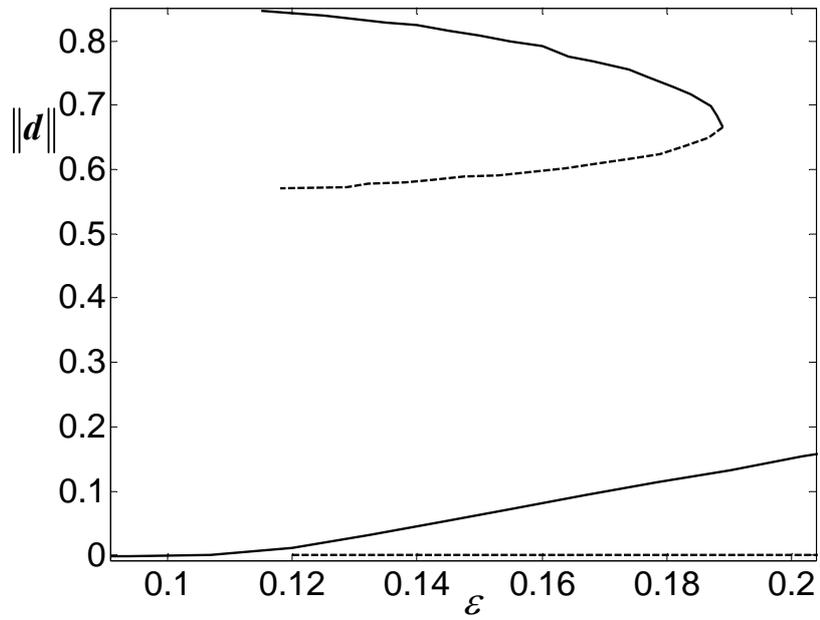

Figure 12

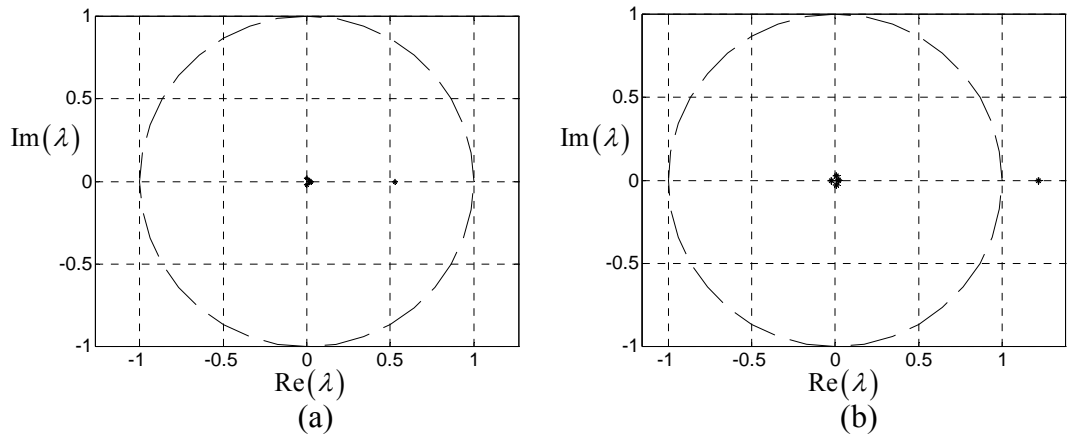

(a) (b)

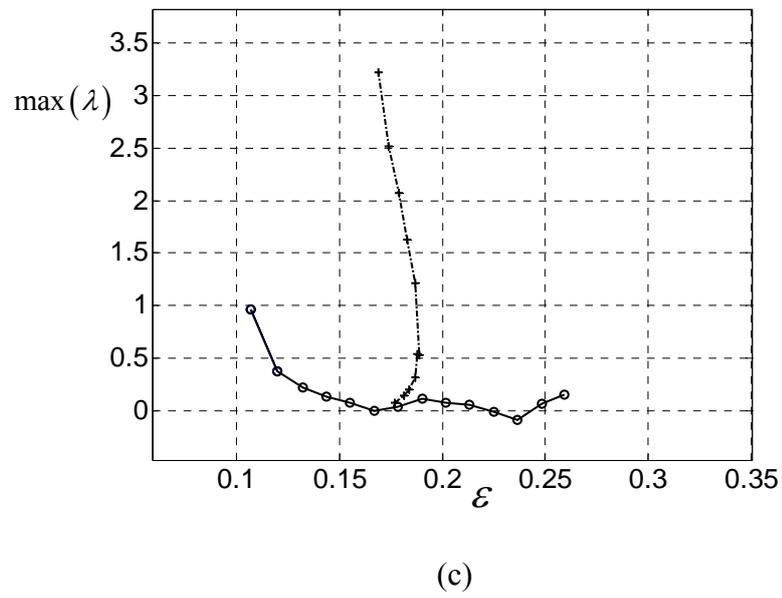

(c)

Figure 13

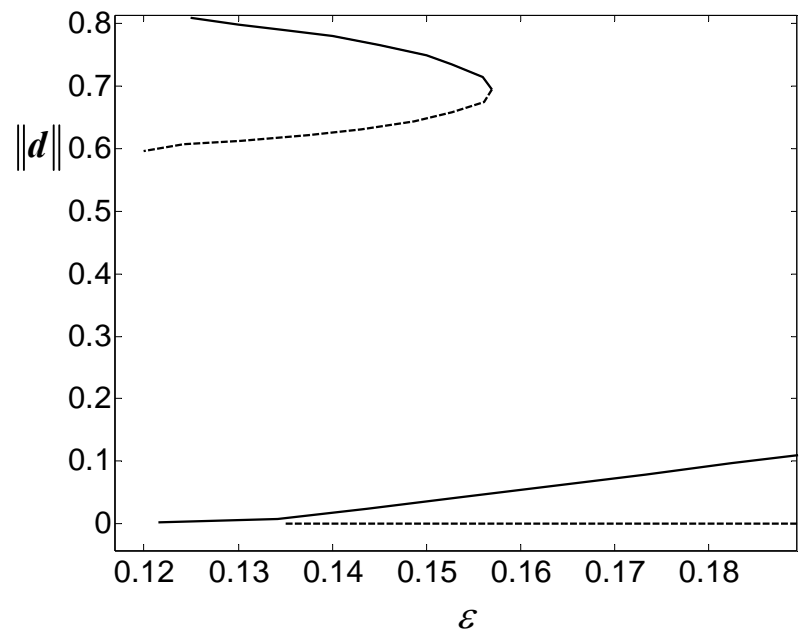

Figure 14

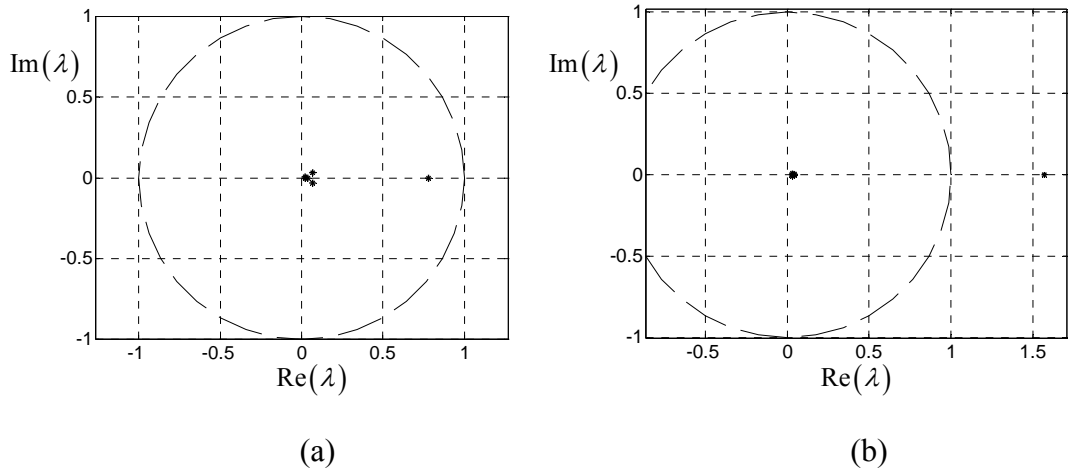

(a)    (b)

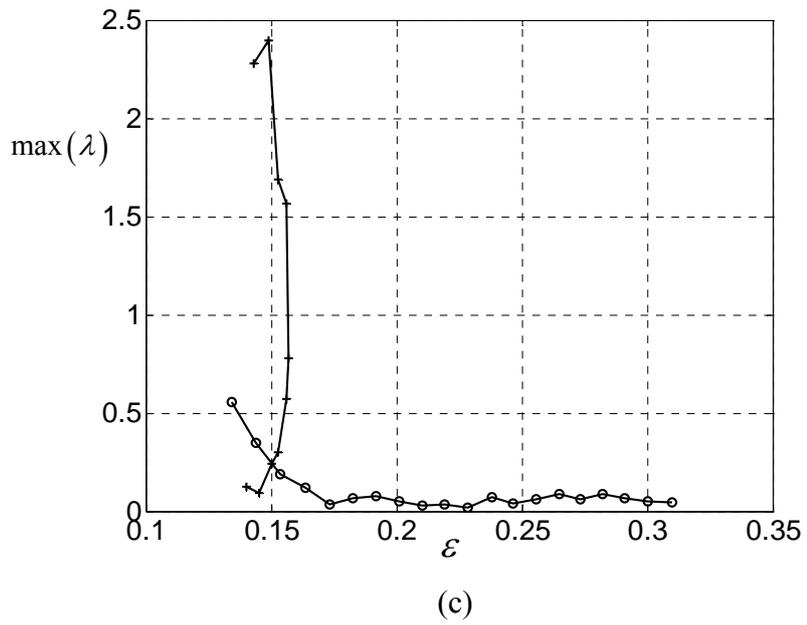

(c)

Figure 15

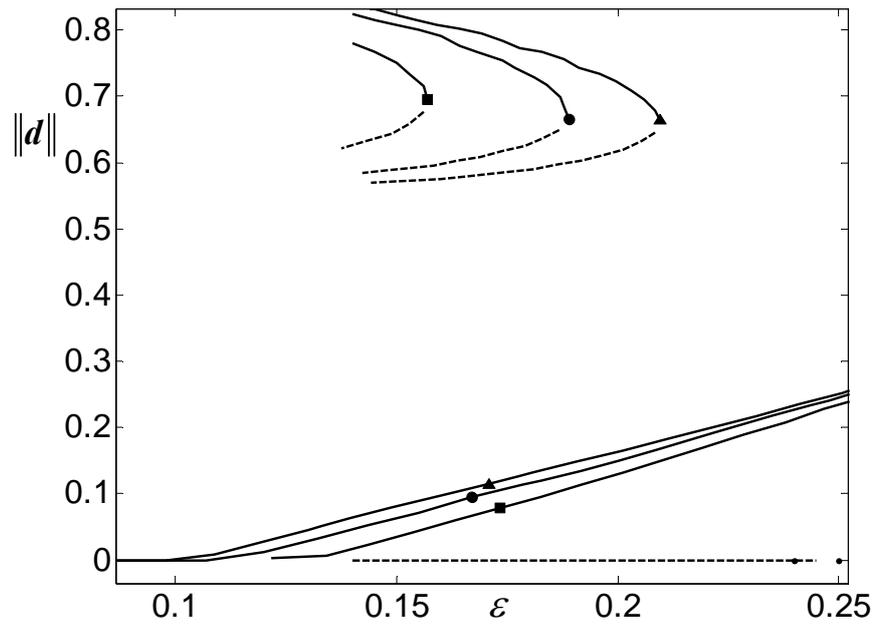

Figure 16

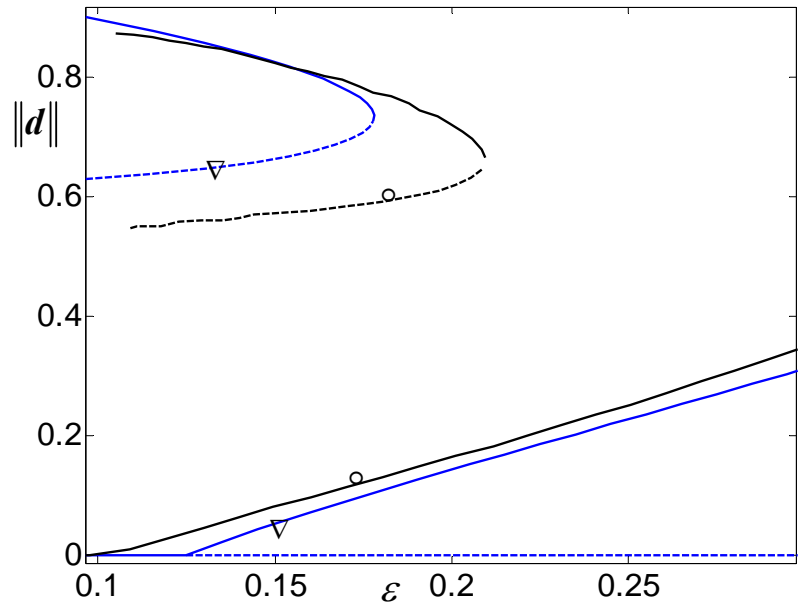

Figure 17

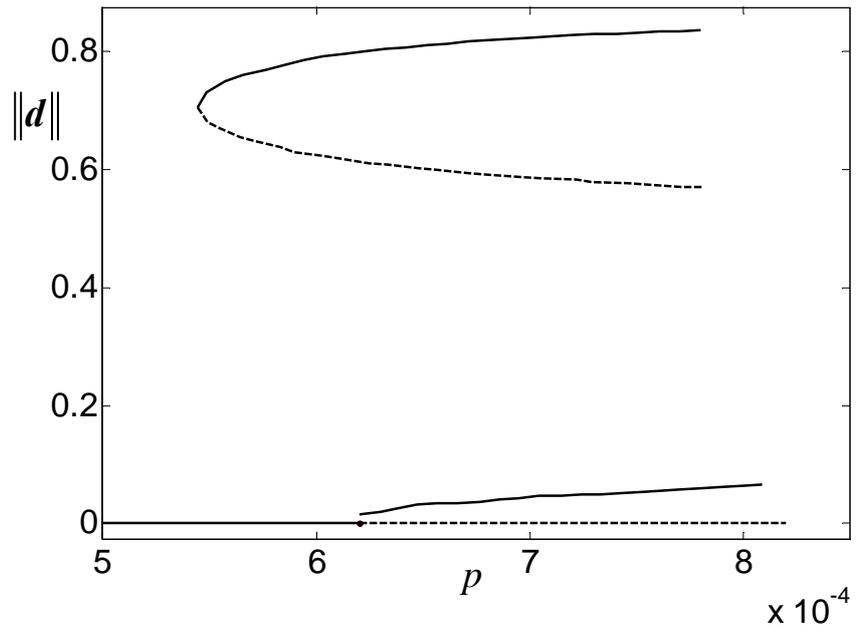

Figure 18

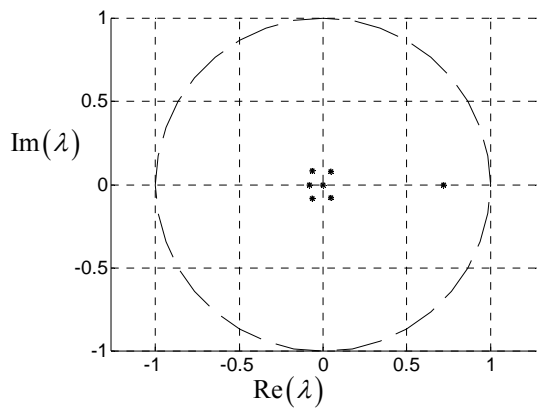 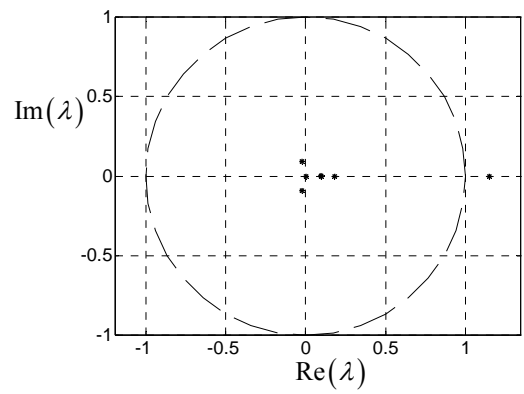

(a)          (b)

Figure 19

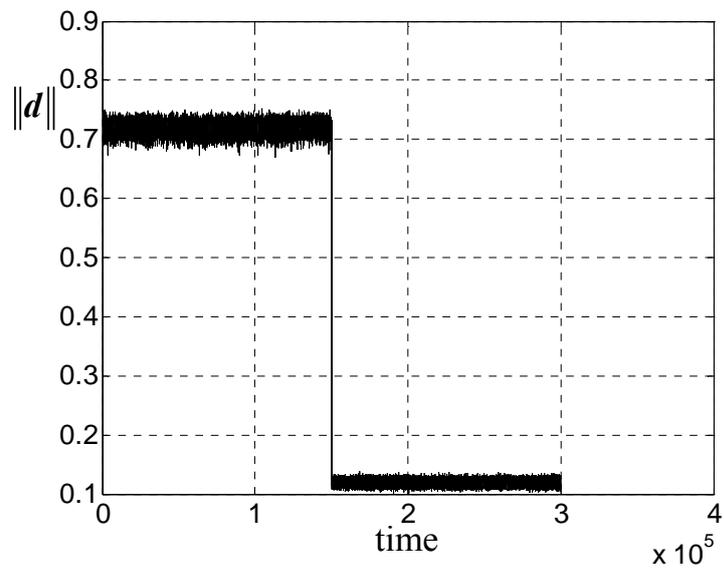

Figure 20

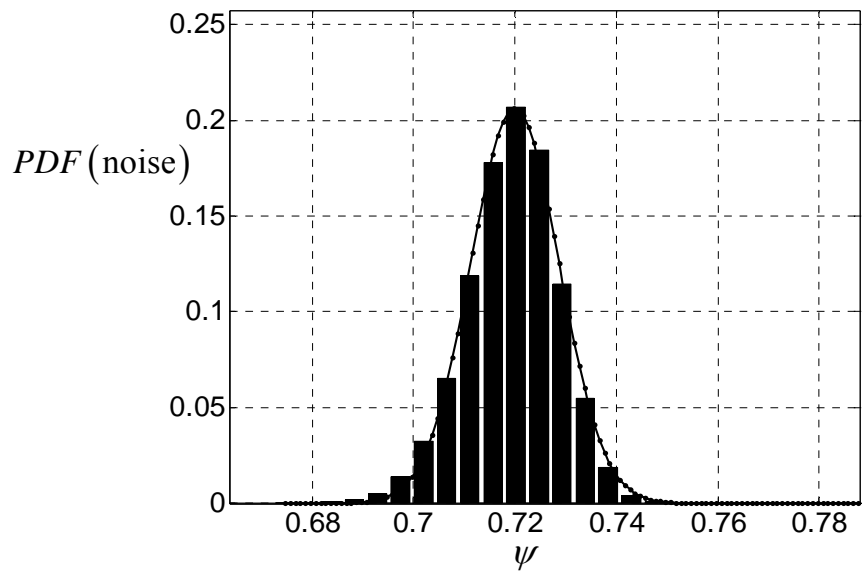

Figure 21

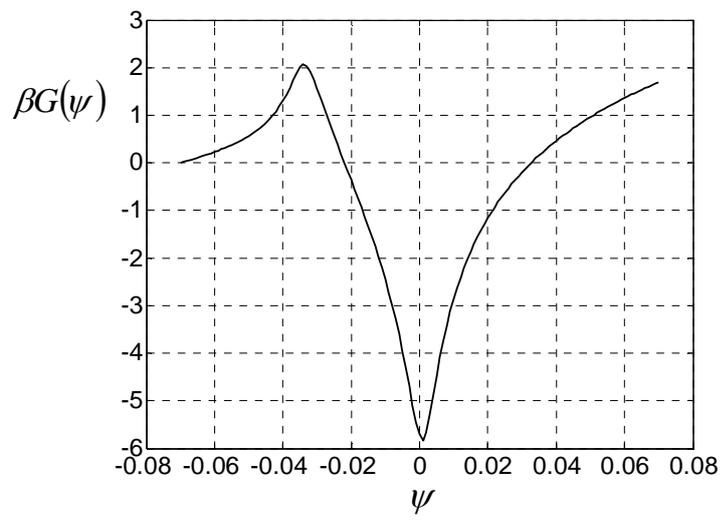

# Captions of Figures

**Figure 1**. The fundamental assumption of the Equation-Free approach: Higher order moments ($u_f$) of an evolving microscopic distribution become "quickly"- within the time-horizon $T$ - functionals of the lower-order moments ($u$).

**Figure 2.** Schematic description of the concept of the Equation-Free approach.

**Figure 3.** Arc-length continuation using the GMRES technique and the timestepper approach.

**Figure 4.** Degree Distribution of an Erdős–Rényi network constructed with $p = 0.0008$ (o) compared with the degree distribution of a network of 10000 individuals which follows the Poisson distribution (13) with a parameter $\bar{k} = N \cdot p = 10000 \cdot 0.0008 = 8$ (*).

**Figure 5.** An example of the evolution rules. The open circle represents an inactive neuron, while the filled circle represents un activated neuron (**a**) $p_{0 \to 1} = 1 - \varepsilon$, $\sigma_i(t) = \sum_{j \in \Lambda_i} a_j(t) = \sum_{j=1}^{9} a_j(t) = 6 > \deg ree(i)/2$. Here $\deg ree(i) = 9$. (**b**) Similar, $p_{1 \to 0} = 1 - \varepsilon$ $\sigma_i(t) = \sum_{j \in \Lambda_i} a_j(t) = \sum_{j=1}^{5} a_j(t) = 1 < \deg ree(i)/2$, $\deg ree(i) = 5$.

**Figure 6:** Temporal evolution of the norm of densities for an Erdős – Rényi network constructed with a connectivity probability $p = 0.0006$ for different values of $\varepsilon$, starting from different initial conditions for probability (**a**) $\varepsilon = 0.1$ (**b**) $\varepsilon = 0.14$ (**c**) $\varepsilon = 0.16$ (**d**) $\varepsilon = 0.2$.

**Figure 7:** Temporal evolution of the norm of densities of degrees with connectivity probability $p = 0.0007$ (**a**) $\varepsilon = 0.1$ (**b**) $\varepsilon = 0.14$ (**c**) $\varepsilon = 0.16$ (**d**) $\varepsilon = 0.2$.

**Figure 8**: Temporal evolution of the norm of densities, for different values of probability $\varepsilon$ starting from different initial conditions constructed with connectivity probability $p = 0.0008$ (**a**) $\varepsilon = 0.1$ (**b**) $\varepsilon = 0.14$ (**c**) $\varepsilon = 0.16$ (**d**) $\varepsilon = 0.2$.

**Figure 9**: Coarse-grained bifurcation diagram with respect to the activation probability $\varepsilon$, for an Erdős–Rényi network constructed with connectivity probability $p = 0.0008$. Solid lines correspond to the coarse-grained stable states while the dotted lines correspond to unstable ones. There is a saddle-node and a transcritical bifurcation.

**Figure 10**: Six largest eigenvalues around the turning point, for an Erdős–Rényi network constructed with connectivity probability $p = 0.0008$ as approximated using the Arnoldi eigensolvder: (**a**) on the stable and (**b**) on the unstable branch, (**c**) The leading eigenvalue of the linearized coarse-grained dynamics vs. the bifurcation parameter $\varepsilon$.

**Figure 11**: Coarse-grained bifurcation diagram with respect to the activation probability $\varepsilon$, for an Erdős–Rényi network constructed with connectivity probability $p = 0.0007$. Solid lines correspond to the coarse-grained stable states while the dotted lines correspond to unstable ones. There is a saddle-node and a transcritical bifurcation.

**Figure 12**: Six largest eigenvalues around the turning point, for an Erdős–Rényi network constructed with connectivity probability $p = 0.0007$ as approximated using the Arnoldi eigensolver: **(a)** on the stable and **(b)** on the unstable branch, **(c)** The leading eigenvalue of the linearized coarse-grained dynamics vs. the bifurcation parameter $\varepsilon$.

**Figure 13**: Coarse-grained bifurcation diagram with respect to the activation probability $\varepsilon$, for an Erdős–Rényi network constructed with connectivity probability $p = 0.0006$. Solid lines correspond to the coarse-grained stable states while the dotted lines correspond to unstable ones. There is a saddle-node and a transcritical bifurcation.

**Figure 14**: Six largest eigenvalues around the turning point, for an Erdős–Rényi network constructed with connectivity probability $p = 0.0006$ as approximated using the Arnoldi eigensolver: **(a)** on the stable and **(b)** on the unstable branch, **(c)** The leading eigenvalue of the linearized coarse-grained dynamics vs. the bifurcation parameter $\varepsilon$.

**Figure 15**: Bifurcation diagrams with respect to the activation probability $\varepsilon$, for Erdős–Rényi networks constructed with connectivity probability $p = 0.0008$ (squares) $p = 0.0007$ (circles) $p = 0.0006$ (triangles).

**Figure 16**: Bifurcation diagrams in the case of the "mean-field" approximation (marked with a triangle) compared to the Erdős – Rényi type network with $p = 0.0008$ (marked with a circle).

**Figure 17**: Bifurcation diagram with respect to the connection probability $p$ (in the superset of all Erdős–Rényi type networks) for a constant value of $\varepsilon$ ($\varepsilon = 0.14$). Solid lines correspond to the coarse-grained stable states while the dotted line correspond to the unstable ones.

**Figure 18**: Six largest eigenvalues around the turning point as approximated using the Arnoldi eigensolver **(a)** on the stable and **(b)** on the unstable branch.

**Figure 19**: Long-run temporal simulation of a microscopic realization for and Erdős–Rényi network constructed with connectivity probability $p = 0.0007$ and $\varepsilon = 0.181$.

**Figure 20**: Probability density estimate of the noise for $p = 0.0007$ and $\varepsilon = 0.181$. Bars correspond to the probability density function of the noise as computed from a long run simulation for $t = 40000$ time steps. The solid line corresponds to the probability density function of a Gaussian distribution with a mean value $m = 0.72$ and standard deviation $\sigma = 0.0123$.

**Figure 21**: The coarse-grained free energy for $p = 0.0007$ and $\varepsilon = 0.181$ with respect to the reaction coordinate $\psi$ as defined in Eq. (15). The minimum of the curve corresponds to the coarse-grained metastable equilibrium and the maximum to the unstable one.